\documentclass[10pt, a4paper, twocolumn]{article} 

\usepackage[english]{babel} 

\usepackage{microtype} 

\usepackage{amsmath,amsfonts,amsthm} 

\usepackage[svgnames]{xcolor} 

\usepackage[hang, small, labelfont=bf, up, textfont=it]{caption} 

\usepackage{booktabs} 

\usepackage{lastpage} 

\usepackage{graphicx} 

\usepackage{enumitem} 
\setlist{noitemsep} 

\usepackage{sectsty} 
\allsectionsfont{\usefont{OT1}{phv}{b}{n}} 

\usepackage{aas_macros}  

\usepackage{geometry} 

\usepackage[T1]{fontenc} 
\usepackage[utf8]{inputenc} 

\usepackage{XCharter} 

\usepackage{fancyhdr} 
\fancypagestyle{firstpage}{ 
	\fancyhf{}
}

\newcommand{\authorstyle}[1]{{\large\usefont{OT1}{phv}{b}{n}\color{DarkRed}#1}} 

\usepackage{titling} 

\newcommand{\HorRule}{\color{DarkGoldenrod}\rule{\linewidth}{1pt}} 

\pretitle{
	\vspace{-30pt} 
	\HorRule\vspace{10pt} 
	\fontsize{32}{36}\usefont{OT1}{phv}{b}{n}\selectfont 
	\color{DarkRed} 
}

\posttitle{\par\vskip 15pt} 

\preauthor{} 

\postauthor{ 
	\vspace{10pt} 
	\par\HorRule 
	\vspace{20pt} 
}

\usepackage{lettrine} 
\usepackage{fix-cm}	

\newcommand{\initial}[1]{ 
	\lettrine[lines=3,findent=4pt,nindent=0pt]{
		\color{DarkGoldenrod}
		{#1}
	}{}%
}

\usepackage{xstring} 

\newcommand{\lettrineabstract}[1]{
	\StrLeft{#1}{1}[\firstletter] 
	\initial{\firstletter}\textbf{\StrGobbleLeft{#1}{1}} 
}

\usepackage[backend=bibtex,style=authoryear,natbib=true]{biblatex} 

\usepackage[autostyle=true]{csquotes} 

\title{Uranus and Neptune are key to understand planets with hydrogen atmospheres} 

\author{
	\authorstyle{Tristan Guillot} 
	\newline\newline 
	Universit\'e C\^ote d'Azur, Laboratoire Lagrange, OCA, CNRS UMR 7293, Nice, France 
}

\date{A White Paper for ESA's Voyage 2050 --- \today} 

\begin{document}

\maketitle 

\thispagestyle{firstpage} 


\lettrineabstract{Uranus and Neptune are the last unexplored planets of the Solar System. I show that they hold crucial keys to understand the atmospheric dynamics and structure of planets with hydrogen atmospheres. Their atmospheres are active and storms are believed to be fueled by methane condensation which is both extremely abundant and occurs at low optical depth. This means that mapping temperature and methane abundance as a function of position and depth will inform us on how convection organizes in an atmosphere with no surface and condensates that are heavier than the surrounding air, a general feature of gas giants. Using this information will be essential to constrain the interior structure of Uranus and Neptune themselves, but also of Jupiter, Saturn and numerous exoplanets with hydrogen atmospheres. Owing to the spatial and temporal variability of these atmospheres, an orbiter is required. A probe would provide a reference profile to lift ambiguities inherent to remote observations. It would also measure abundances of noble gases which can be used to reconstruct the history of planet formation in the Solar System. Finally, mapping the planets' gravity and magnetic fields will be essential to constrain their global composition, structure and evolution.    
}


\section{Introduction}
 
Thus far, mankind has put spacecrafts in orbit around six planets. Mercury has been visited by Mariner 10 and MESSENGER and is awaiting BepiColombo. Venus was visited successfully by a score of space missions including the recent Venus Express and Akatsuki. The Earth has had satellites since Sputnik 1 in 1957. They are now countless and our night sky is now threatened by a float of thousands bright low-orbit satellites for internet communication. Mars is currently orbited by six still operational spacecrafts, including ExoMars TGO. 
Jupiter was orbited by Galileo between 1995 and 2003, by Juno since 2016 and will be by JUICE starting in 2029. Saturn has had Cassini between 2004 and 2017. The two remaining planets in the Solar System, Uranus and Neptune, have only been visited for a couple of days each and from a distance by the Voyager 2 spacecraft, never by orbiters. 

\begin{figure*}
	\includegraphics[width=\linewidth]{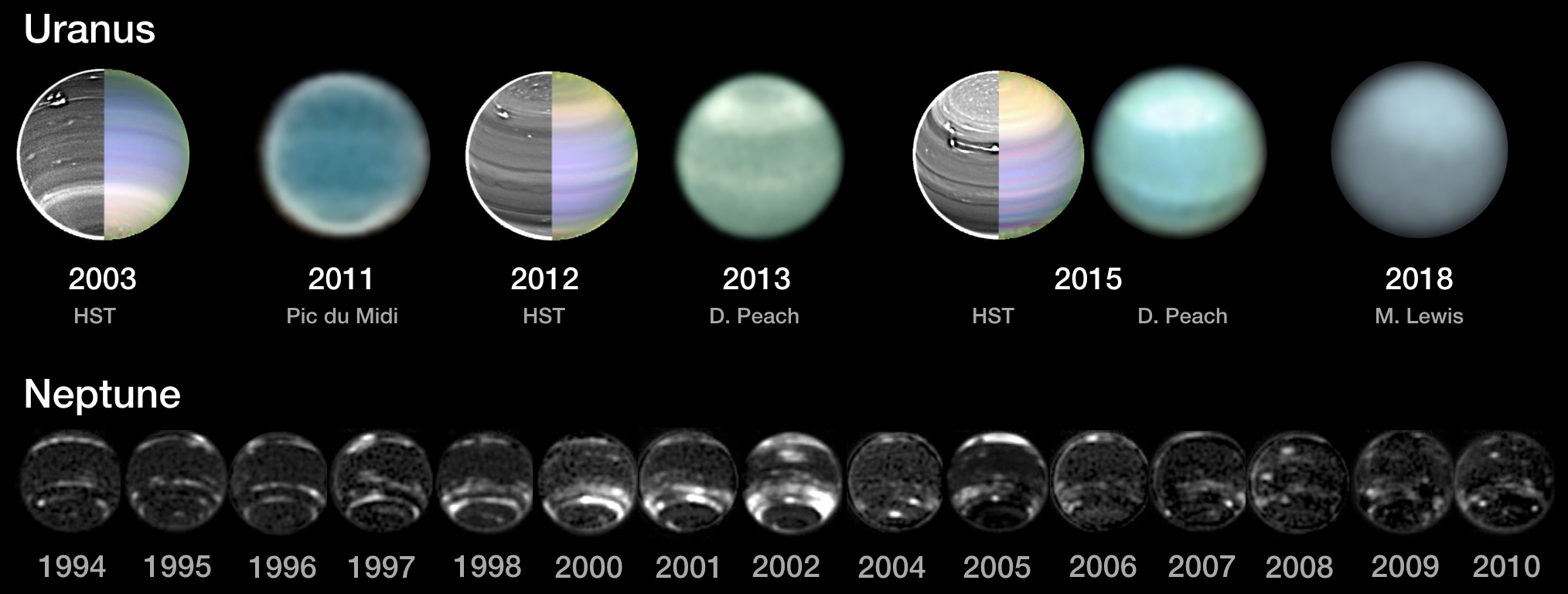} 
\caption{Images of Uranus and Neptune showing seasons and storms. The HST/STIS images of Uranus correspond to H band (left) and false color (right) images \citep{Sromovsky+2019}. Amateur images from the Pic du Midi, D. Peach and M. Lewis have been taken from the PVOL database (http://pvol2.ehu.eus/). The images of Neptune have been obtained from HST/WFPC2 in the visible \citep{Karkoschka2011b}.}
	\label{fig:globes} 
\end{figure*}

Yet, both Uranus and Neptune are fascinating planets that hold some of the keys to understand the origin of our Solar System and to make sense of the observations of exoplanetary atmospheres. As seen in Fig.~\ref{fig:globes}, they both have active, complex atmospheres, observed and monitored by professional and amateurs alike.  I will advocate that the exploration of our Solar System must continue and that either Uranus or Neptune, or both, should be the next targets in this journey. 

I will take a subjective and admittedly biased approach: I will focus on the giant planets themselves rather than on their moons, their rings or their complex magnetospheres. I will first review lessons from missions at Jupiter and Saturn, discuss the importance of the methane cloud layer, review present knowledge concerning Uranus' and Neptune's interior structure and composition and then derive mission objectives.

\section{Lessons from Galileo, Juno and Cassini}

Leaving the Sun apart, giant planets hold most of the mass of the Solar System. Yet their composition remains poorly constrained. This greatly hinders our ability to reconstruct the history of the Solar System. With Juno and Cassini, we did make great progress on our ability to constrain the interior compositions, dynamics and magnetic fields of Jupiter and Saturn. However, both their interiors and atmospheres appear more complex than previously envisioned. Condensing species in the atmosphere appear to have highly variable compositions. The associated deep temperature structure is unknown. This limits our understanding of the interiors of these planets and the constraints that we can derive on their composition. On the other hand, Uranus and Neptune hold some of the keys to understand how convection organizes in these atmospheres and therefore to better understand giant planets as a whole. 

Two major particularities of the atmospheres of giant planets are the absence of a surface and the fact that condensates are heavier than surrounding air, creating a meteorological regime that is intrinsically different from that of terrestrial planets. In the Earth atmosphere, dry air has a mean molar mass of about 29, compared to 18 for water. This means that moist air naturally tends to rise, slowly if the relative humidity is less than 100\%, much more rapidly during storms due to water condensation and latent heat release. In giant planets, and generally in planets with hydrogen-helium atmospheres, the mean molar mass of ``air'' is much smaller, about 2.3 for a solar composition hydrogen-helium mixture. This means that moist air then tends to sink. Since there is no surface, it is not clear where condensing species will sink to. In spite of this, since these planets are convective and storms are regularly observed, the prevailing view has been that this is a minor effect that can be largely ignored: Convective motions should homogenize composition below the condensation level (the ``cloud base'') and latent heat effects should lead to powerful storms capable of an efficient upward transport of condensable species. The Galileo probe measurements \citep{Wong+2004} and the Juno measurements \citep{Bolton+2017, Li+2017} have shown that this view is at best incomplete and perhaps altogether wrong. 

Prior to 1995, the conventional view was that Jupiter should at least have a solar abundance of water, that condensation (clouds and rainout) would limit its abundance at pressures smaller than 5 bars \citep{Weidenschilling+Lewis1973,Atreya+1999}, but that in the deep atmosphere we should reach a uniform abundance of water corresponding to the 'bulk' abundance.  The Galileo probe reached much deeper than 5 bars: in fact it went down to 22 bars and a temperature of 425\,K \citep{Seiff+1998}. Surprisingly, the water abundance that was still significantly subsolar and still rising at that level \citep{Wong+2004}. The explanation has been that the Galileo probe had fallen into a particular region of Jupiter's atmosphere, a hot spot, location of a significant downdraft due to a Rossby wave circling the planet \citep{Showman+Ingersoll1998}: We had been unlucky. Surely, if the Galileo probe had been sent elsewhere than in one of these hot spots (covering less than 1\% of the surface of the planet), we should have experienced a more 'normal' situation. 

About twenty years after the Galileo probe, the Juno MWR instrument measured the ammonia abundance in Jupiter's atmosphere and  found yet another puzzling situation: Ammonia, which is condensing at lower temperatures (around 150K and pressures of ~0.7 bar in Jupiter) was found to show a very non-uniform abundance as a function of depth, down to pressure levels of at least 30 bars \citep{Li+2017}, i.e. 4 pressure scale heights deeper than the ammonia cloud base! Furthermore, the abundance is also highly latitudinally variable, with the equatorial zone presenting a high ammonia abundance that is rather uniform with depth while the other latitudes show lower values, confirming ground based observations \citep{dePater+2016}. This is surprising and cannot be explained by meridional circulation \citep{Ingersoll+2017}. Instead, it is believed that apart water storms are able to penetrate high-enough into the planet's atmosphere to create ammonia-loaded hailstones ('mushballs') that deplete the upper atmosphere of its ammonia (Guillot et al., submitted). The absence of storms and lightning at the equator \citep{Brown+2018}  is consistent with the high abundance of ammonia there. The model also implies that the water and temperature field should be non-uniform latitudinally and vertically, and that it should be time-variable.   

Observations in Saturn point to the same process in Saturn: The abundance of ammonia inferred from 5 microns observations from Cassini VIMS is high at the equator and low at other latitudes \citep{Fletcher+2011}. Furthermore, the recurrence of giant storms in the atmosphere has been shown to be linked to water storms and the recurrent progressive cooling of Saturn's atmosphere and its heating after a storm \citep{Li+Ingersoll2015}. 

It therefore appears that local storms play an important, perhaps dominant role in controlling the structure of the atmosphere. After all, it has been estimated from Galileo observations that storms can transport the entire heat flux of Jupiter \citep{Gierasch+2000}. Of course, meridional circulation plays a role \citep[see][]{Fletcher+2019}, but it could dominate mostly in the stratosphere and upper troposphere, and play a minor role in the deep troposphere and interior of the planet. This implies that the distribution of all condensing species (methane in Uranus and Neptune, ammonia, hydrogen sulfur, water, silicates...) are in question. So are the temperature profiles in these regions. This even can be extended to include helium, which separates from metallic hydrogen at pressures around 1 Mbar \citep{Stevenson+Salpeter1977}. 

The fact that we fail to understand such basic elements has profound implications for our knowledge of giant planets in general. It means that the atmospheric boundary condition used for interior models and the internal adiabat used are perhaps not well-defined, thus raising doubts (or at least increasing the uncertainties) of constraints derived from interior models \citep[e.g.,][]{Guillot2005}. It will also limit what we can interpret from observations of exoplanets with hydrogen atmospheres. 

Understanding how hydrogen atmospheres transport heat and elements is a formidable task. It involves multiple scales, from the global scale (i.e., the size of the planet itself, $\sim 100,000$\,km) to the sizes of storms ($\sim 1-100$ km) and includes complex hydrodynamics and microphysics. Global circulation models \citep[e.g.][]{Dowling+1998, Kaspi+2009, Liu+Schneider2010, Guerlet+2014} must simplify the treatment of storms and clouds. Cloud or cloud-ensemble models \citep[e.g.][]{Hueso+SanchezLavega2001, Sugiyama+2014, Li+Chen2019} do not include meridional motions and/or global scale winds. They also simplify the microphysics. Detailed microphysical treatments \citep[e.g.][]{Yair+1995} are based on the Earth's schemes and must be extrapolated to be applied to the giant planets. Therefore, numerical simulations can only guide us on what may be occurring in these atmospheres. We need ground truth. 

Unfortunately, the measurements required to validate models and understand what is going on are scarce because in Jupiter and Saturn most of the action occurs hidden from view at large optical depth. The structure of the ammonia condensation region near 0.7 bar in Jupiter and 1.5 bar in Saturn is observable, but ammonia has a low abundance ($\sim 100$ to $500$\,ppmv mixing ratio) and can only drive a weak moist convection \citep[e.g.][]{Stoker1986}. Instead, most of the storms that we see must be powered by water condensation \citep[see][]{Hueso+SanchezLavega2001, Hueso+2002, Sugiyama+2014, Li+Chen2019} , at levels of $\sim 6$\,bar in Jupiter and $\sim 12$\,bar in Saturn. Juno's MWR instrument was able to probe these regions and deeper in Jupiter but the measurements are mostly sensitive to ammonia's absorption, now believed to be a complex function of depth, latitude and possibly even longitude \citep{Li+2017}. The effect of water is indirect. Finally, we lack a well-defined temperature pressure profile that would allow lifting some of the degeneracies in the measurements. 

Uranus and Neptune possess one key ingredient to understand atmospheric dynamics in hydrogen atmospheres: They are cold enough for methane to condense at low pressure levels ($\sim 1.5$\,bar) in a region of the troposphere at modest optical depth, and methane is present in abundance to drive moist convection at these levels \citep{Stoker+Toon1989}.

\section{Probing the methane condensation region in Uranus and Neptune} 

In Uranus and Neptune, radio occultations from Voyager 2 indicate that methane condensation should occur at pressures of around 1.5 bar for a temperature of about 80\,K \citep{Lindal1992}. Methane is extremely abundant and, as observed for ammonia in Jupiter and Saturn, its abundance is variable with latitude. The maximum mixing ratio in Uranus inferred from HST, Keck and IRTF observations is $f_{\rm CH_4}=2.55\%$ to $3.98\%$ \citep{Sromovsky+2019}, corresponding to a mixing ratio by mass in the range $q_{\rm CH_4}=0.154-0.224$. In Neptune, the maximum value detected with VLT/Muse at a latitude $30^\circ$S is even higher, $f_{\rm CH_4}=5.90\pm 1.07\%$ \citep{Irwin+2019a}, corresponding to $q_{\rm CH_4}=0.30\pm 0.04\%$. Thus, for both planets, methane accounts for 15\% to 30\% of the mass in the upper atmosphere. For comparison, ammonia in Jupiter and Saturn represents only about 0.3\% of the atmospheric mean molecular weight. 

This means that, in Uranus and Neptune, the atmospheric mean molecular weight increases from $\mu\sim 2.3$ at $P<1\,$bar to $\mu\sim 2.6-3.1$ at greater depth, where methane has reached its bulk abundance. This is a considerable increase yielding a highly stabilizing, bottom-heavy atmosphere. In fact, the abundance of methane in both planets even exceeds the critical value over which moist convection is inhibited \citep{Guillot1995}. It has been shown that this inhibition also extends to double-diffusive convection and could yield a highly super-adiabatic temperature gradient \citep{Leconte+2017,Friedson+Gonzales2017}. 

It is important to recognize at this point that the temperature profile inferred from the Voyager radio-occultations is highly degenerate. The physical quantity that is measured is the refractivity as a function of height. The refractivity depends on both the mean molecular weight and the temperature. The profile widely used for Uranus and Neptune \citep{Lindal1992} corresponds to one possible choice, but other solutions are possible \citep{Guillot1995, Sromovsky+2011}. Direct measurements of temperature profiles at several locations in Uranus or Neptune's atmosphere, down to several bars is crucial. 

\begin{table*}[]
\caption{Parameters characterizing the main cloud layers in Jupiter, Saturn, Uranus and Neptune. $f$ is the volume mixing ratio of the condensing species, $P_{\rm ref}$ and $T_{\rm ref}$ the reference pressure and temperature at cloud base, $\Delta T_L$, $\Delta T_\mu$ and $\xi_{\rm inhib}$ are defined in the text.}\label{tab:clouds}
\begin{tabular}{lcccccc}\hline\hline
Condensation layer & $f\rm\ [\%]$ & $P_{\rm ref}\rm\ [bar]$ & $T_{\rm ref}\rm\ [K]$ & $\Delta T_L\rm\ [K]$ & $\Delta T_\mu\rm\ [K]$ & $\xi_{\rm inhib}$ \\ \hline
NH$_3$ in Jupiter & $0.03-0.04$ & 0.7 & 150 & $0.29-0.35$ & $0.31-0.38$ & $0.040-0.048$ \\
NH$_3$ in Saturn & $0.03-0.05$ & 1.5 & 150 & $0.27-0.44$ & $0.29-0.48$ & $0.036-0.061$ \\
H$_2$O in Jupiter$^*$ & $0.2-0.6$ & 6 & 300 & $2.7-7.9$ & $4.3 -12.7$ & $0.23-0.65$ \\
H$_2$O in Saturn$^*$ & $0.4-1.2$ & 12 & 300 & $5.4-15.2$ & $8.6-25.4$ & $0.45-1.25$\\ \hline
CH$_4$ in Uranus & $2.5-4.0$ & 1.5 & 80 & $7.6-11.0$ & $12.1-19.0$ & $1.6-2.3$ \\
CH$_4$ in Neptune & $4.8-7.0$ & 1.5 & 80 & $12.8-16.9$ & $23.0-33.2$ & $2.6-3.5$ \\
H$_2$S in Uranus & $0.03-0.08$ & 7 & 120 & $0.21-0.64$ & $0.36-1.1$ & $0.06-0.18$ \\
H$_2$S in Neptune & $0.03-0.08$ & 7 & 120 & $0.21-0.64$ & $0.36-1.1$ & $0.06-0.18$ \\
\hline\hline
\multicolumn{7}{l}{$^*$: We assume an O/H enrichment over solar of 2 to 6 times in Jupiter and 4 to 12 times in Saturn.}\\
\end{tabular}
\end{table*}

Let us now define three quantities to characterize possible changes in the temperature profile in condensation regions. 
The first one, $\Delta T_L$, corresponds to changes in temperature due to latent heat release by condensation: Assuming a base dry adiabatic profile, this is the maximum temperature change due to condensation in an upwelling column. (Note that formally, one should consider potential temperature to account for pressure changes, but this simplification is sufficient for our purposes.) We thus write:
\begin{equation}
\Delta T_L \equiv \frac{q_{\rm v} L_{\rm v}}{c_p},
\end{equation}
where $q_{\rm v}$ is the maximum condensate mass mixing ratio, $L_{\rm v}$ is latent heat of vaporization per unit mass and $c_p$ is the atmospheric heat capacity per mass. 

The second, $\Delta T_\mu$, corresponds to the temperature increase required to compensate for the mean molecular weight change and have a density profile that is neutrally stable to convection, not including possible latent heat effects. Since $\Delta T/T \sim \int d\ln\mu$, it can be shown that:
\begin{equation}
\Delta T_\mu \equiv \left[-\ln\left(1-\varpi q_{\rm v}\right)\right]T,
\end{equation}
where $\varpi\equiv 1-M_{\rm a}/M_{\rm v}$, $M_{\rm a}$ and $M_{\rm v}$ are the molar masses of dry air and the condensing species, respectively, and $T$ is the local temperature in the condensing region. 

The third one is the moist convection inhibition factor $\xi_{\rm inhib}$, defined as \citep{Guillot1995, Leconte+2017}
\begin{equation}
\xi_{\rm inhib}\equiv \frac{\varpi M_{\rm v} L_{\rm v}}{{\cal R} T}q_{\rm v}.
\end{equation}
Moist convection is inhibited whenever $\xi_{\rm inhib}>1$. 

\begin{figure}
	\includegraphics[width=\linewidth]{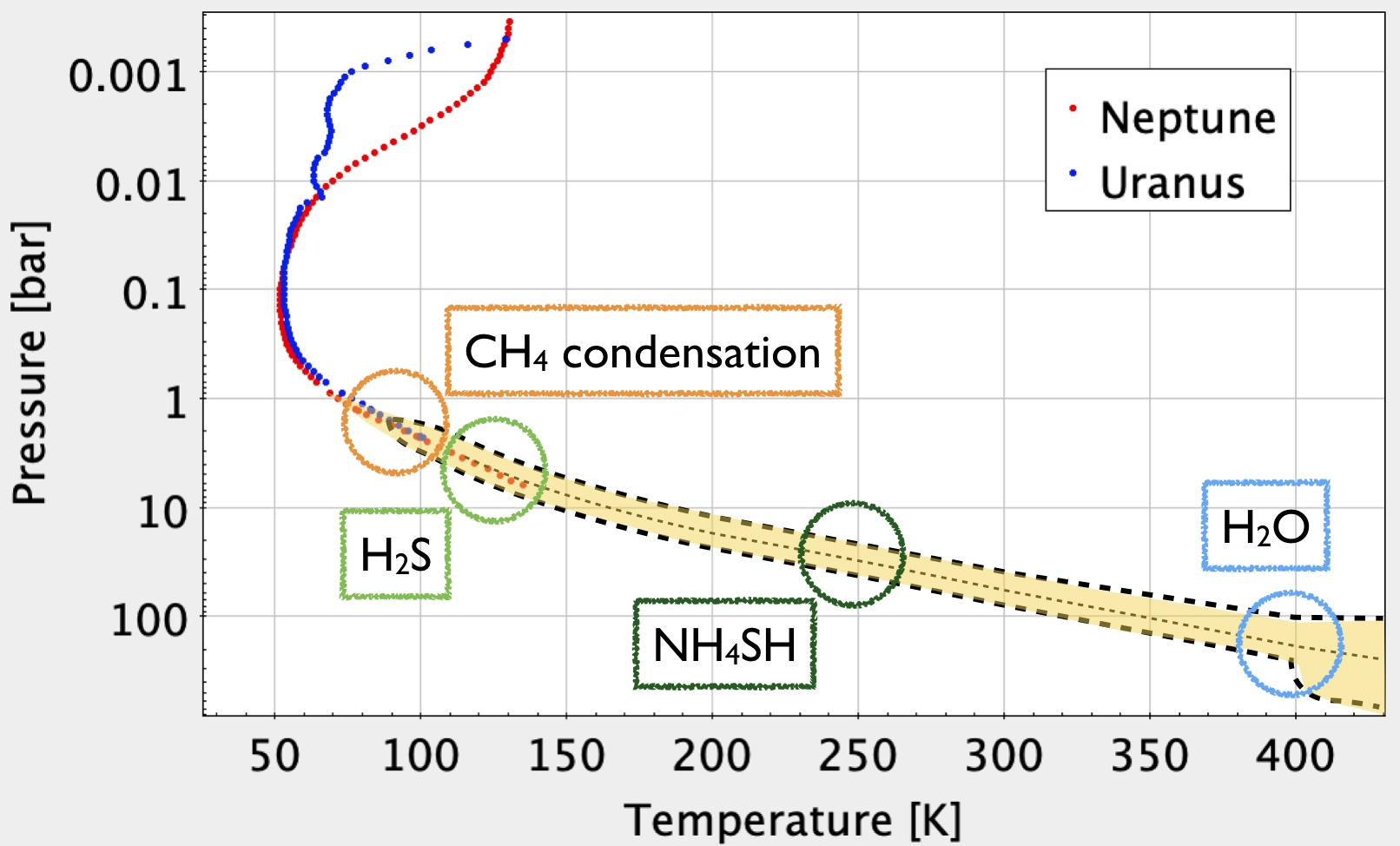} 
	\caption{Temperature pressure profile measured by radio occultation in Uranus (blue dots) and Neptune (red dots) and extended in the deep atmosphere. The yellow area highlights schematically uncertainties on the deep temperature profile. The regions of condensation of the different species are indicated.  } 
	\label{fig:tp} 
\end{figure}

Table~\ref{tab:clouds} lists the main condensation layers potentially accessible to direct observations by a probe or a spacecraft. These can be sorted in two categories: NH$_3$ in Jupiter and Saturn and H$_2$S in Uranus and Neptune are characterized by small values of $\Delta T_L$ and $\Delta T_\mu$. These should lead to weak storms with relatively small updrafts. On the other hand, H$_2$O in Jupiter and Saturn and CH$_4$ in Uranus and Neptune are characterized by large values $\Delta T_L$ and $\Delta T_\mu$ of several Kelvins at least, favorable to the development of large storms with strong updraft velocities. The temperature profile and its uncertainty as envisioned for Uranus and Neptune is depicted in Fig.~\ref{fig:tp}. 

The similarities between H$_2$O in Jupiter and Saturn and CH$_4$ in Uranus and Neptune seen in Table~\ref{tab:clouds} implies that much is to be gained from a detailed characterization of the CH$_4$ condensation layer in Uranus and Neptune. Being at relatively low optical depth, this layer is much easier to characterize than the H$_2$O condensation layer in Jupiter and Saturn, which is at high optical depth and hidden by other thick clouds most of the time. 

\begin{figure*}
	\includegraphics[width=\linewidth]{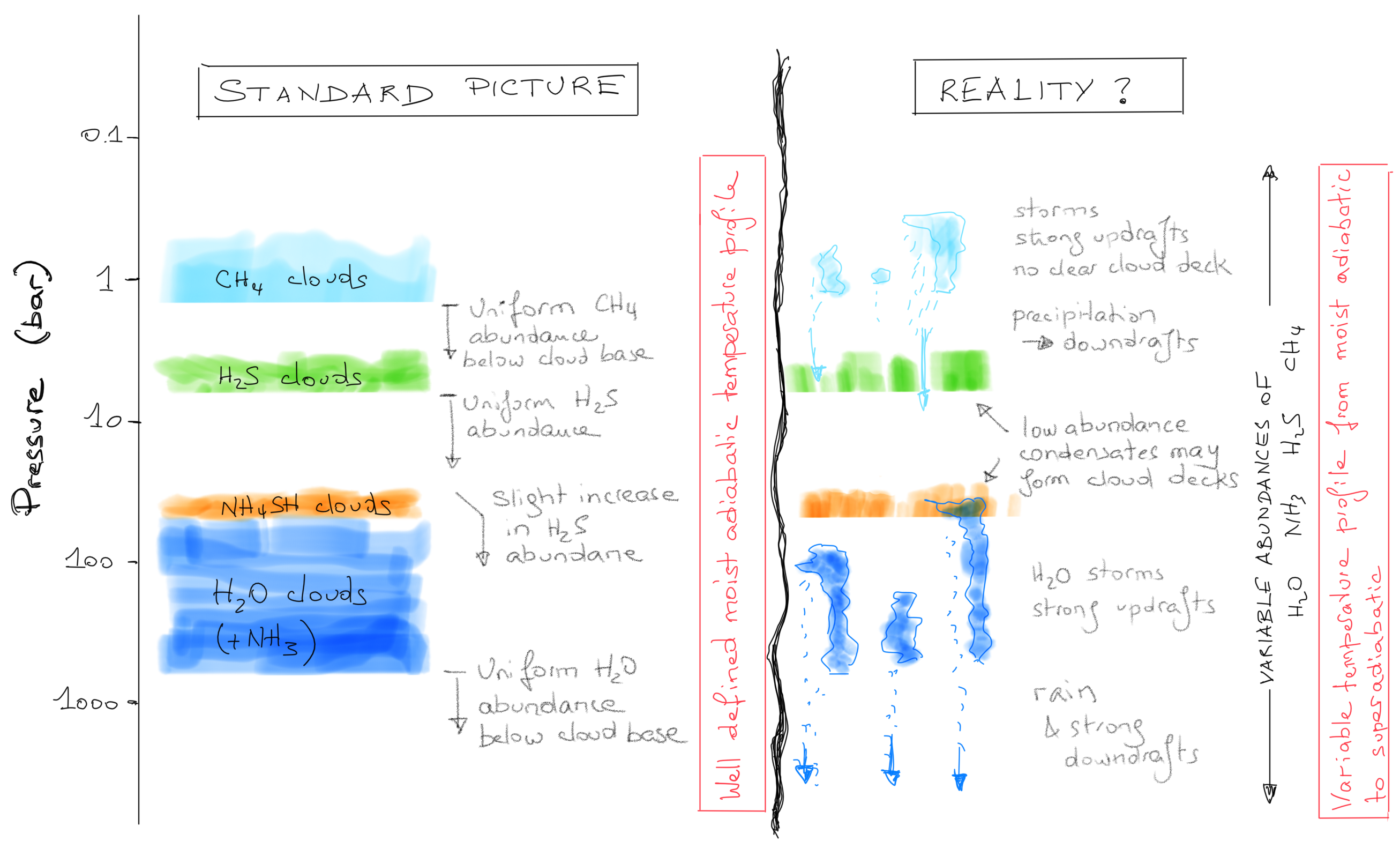} 
	\caption{Sketch of possible cloud structures in Uranus and Neptune. The left side shows the standard picture which assumes that small-scale mixing maintains relatively well-defined cloud decks and a temperature profile close to a moist adiabat (accounting for the condensation of the different species). Any latitudinal variation may be explain by meridional circulation. The right side shows an alternative model in which, for abundant condensing species such as methane and water, storms occur. This implies strong updrafts, but also strong downdrafts due to rainout and evaporative cooling. On the other hand, less abundant species such as H$_2$S and NH$_4$SH may form relatively well-defined cloud decks. In this case, large temperature variations, from moist-adiabatic to super-adiabatic are to be expected (see text).  } 
	\label{fig:clouds} 
\end{figure*}

As sketched in Fig.~\ref{fig:clouds}, condensation in giant planets is generally thought to lead to the formation of relatively well defined cloud layers with an abundance of cloud particles that is essentially a function of the bulk abundance of the condensing species itself. The presence and optical thickness of the cloud decks would then be modulated essentially by other global effects (e.g., being in a zone or belt). For lack of a better alternative, the temperature profile is generally considered close to a moist adiabat. When accounting for storms, the picture may in fact change: Species with large values of $\Delta T_L$ would organize in series of time-variable updrafts, with compensating subsidence drying out the outside environment \citep{Lunine+Hunten1987}. Condensates thus formed would accumulate locally, creating ponds of cold air able to sink much below the cloud base. Thus, instead of consisting in a well-defined cloud deck, we could have instead a relatively clear atmosphere with intermittent storms. Horizontal variations in temperature could be large as indicated by  $\Delta T_L$ and $\Delta T_\mu$. 

This has not been seen in Jupiter and Saturn because the ammonia condensation region is characterized by small values of $\Delta T_L$ and $\Delta T_\mu$ of only a fraction of a Kelvin, much below the intrinsic atmospheric variability and the sensitivity of the measurements. Deeper, NH$_4$SH condensation has an even lower abundance. The water condensation region has these large $\Delta T_L$ and $\Delta T_\mu$ values but is hidden deeper. It is now being probed by Juno/MWR but the temperature variations of a few Kelvins are easily offset by variations in ammonia abundance \citep{Li+2017}, making this determination difficult. 

Variations in ortho-para hydrogen can also contribute to temperature variations, potentially of the same order \citep{Massie+Hunten1982} and have been proposed to lead to layered convection in Uranus and Neptune \citep{Gierasch+Conrath1987}. However the conversion occurs on a timescale measured in years \citep[e.g.][]{Fouchet+2003}, too long to affect directly moist convective events. It may however have an effect on the global circulation of the atmosphere, and it is of course extremely useful to understand global atmospheric motions \citep[e.g.][]{Fletcher+2016}. 

Probing the methane condensation layer (i.e. 1-2 bar, possibly extending measurements to 10 bars) in Uranus and Neptune would give us the ability to decide between the different possibilities of Fig.~\ref{fig:clouds}, understand how heat is transported in hydrogen atmospheres and estimate the entropy in the deep atmosphere. A combination of global and local measurements of temperature and methane abundance is needed.

 \section{The interior structure and composition}
 
 \subsection{Deep boundary condition and latitudinal dependence}
 
The determination of the interior structure and composition of a giant planet relies on accurate determination of its gravitational moments but crucially on theoretical models to describe how density varies with depth in the planet. These models are based on equations of state to reproduce density changes with pressure and temperature, and on a determination of heat transfer in the planet. The atmospheric boundary condition and the internal temperature profile (set by the mechanism responsible for heat transport, i.e., dry, moist or diffusive  convection, radiation and conduction) are essential ingredients in the model \citep[e.g.][]{Guillot2005}. 

\begin{figure}
	\includegraphics[width=\linewidth]{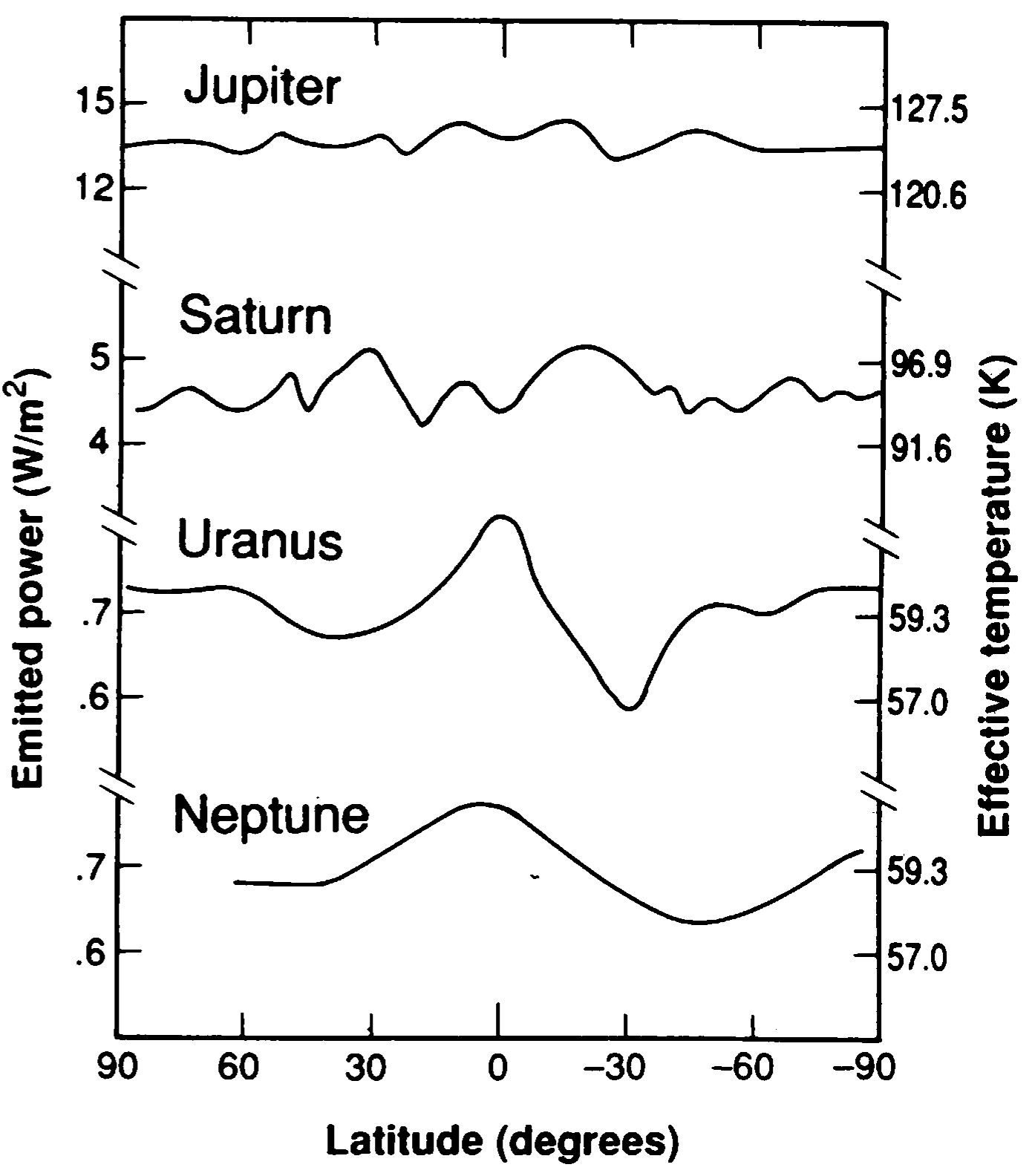} 
	\caption{Emitted infrared flux and equivalent brightness temperature versus latitude for the four outer planets. The radiation is emitted, on average from the 0.3 to 0.5 bar pressure level. \citep[From][]{Ingersoll1990}} 
	\label{fig:brightness} 
\end{figure}

Beyond, a fundamental hypothesis of interior models that is often taken for granted but merits to be revisited is the one-dimensional nature of the problem: All interior and evolution models to date assume that the planetary structure can be solved from a set of one-dimensional differential equations that depend only on an average distance to the center of the planet. This includes of course models for exoplanets \citep[e.g.][]{Guillot2005}. This is comforted by measurements of the infrared flux emitted by Jupiter, Saturn, Uranus and Neptune shown in Fig.~\ref{fig:brightness} \citep{Ingersoll1990, Fletcher+2019} which indicate that, in spite of a highly latitudinally variable insolation, the equator-to-pole temperature variations measured are small ($\pm 2\,$K more or less -compared to about 30\,K on Earth). 
   
Several explanations to account for this constancy of the atmospheric temperature structure are mixing in the atmosphere \citep{Conrath+Gierasch1984}, that thermal gradients at depth counterbalance latitudinal insolation gradients \citep{Ingersoll+Porco1978} or that deep convection generate a heat flux that is stronger at the poles and weaker at the equator \citep{Aurnou+2008}.  

\begin{figure*}
	\includegraphics[width=\linewidth]{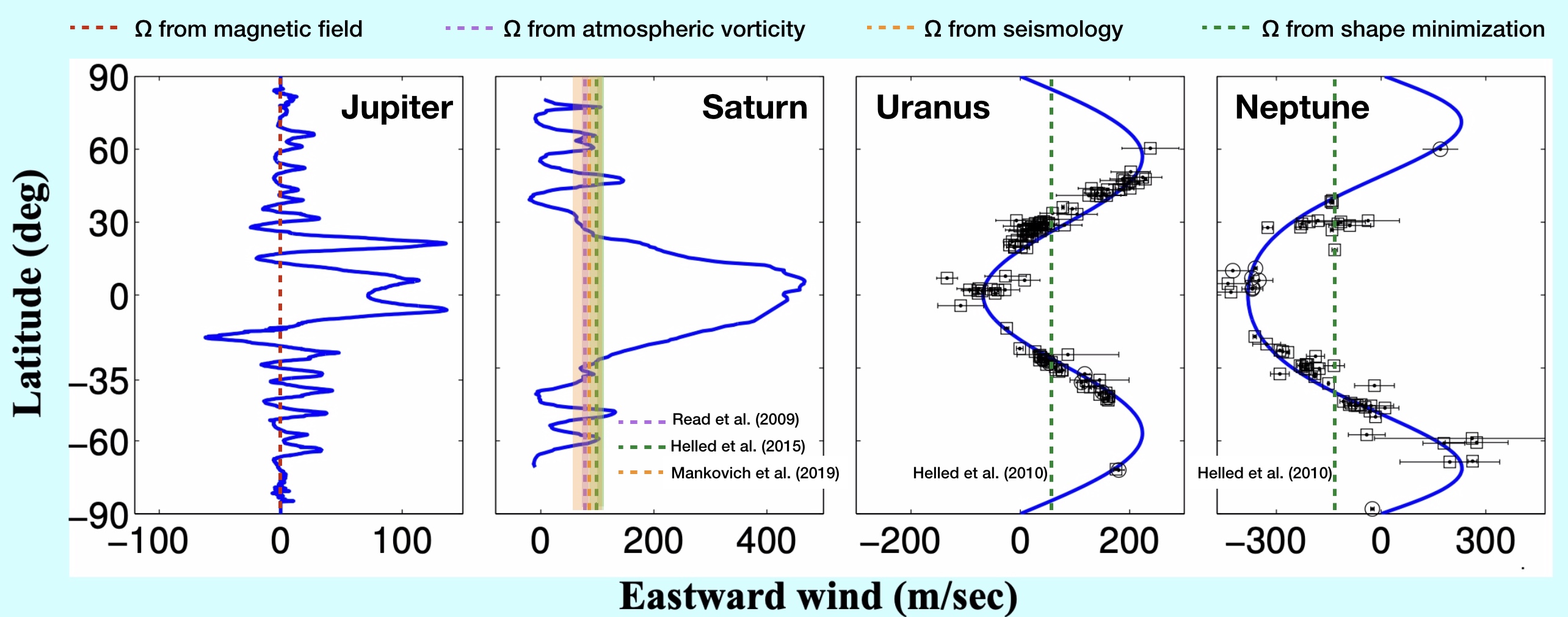} 
	\caption{Zonal winds on the four giant planets as measured by cloud tracking in the so-called system III reference frame \citep[from][and references therein]{Cao+Stevenson2017, Kaspi+2013}. System III is supposed to correspond to the inner magnetic field, but may be accurate only for Jupiter. Other determination of the rotation of the planetary interior have been done by estimating the atmospheric vorticity \citep{Read+2009}, by a method minimizing the difference between the observed and theoretical planetary shape \citep{Helled+2010, Helled+2015}, and by planetary seismology \citep{Mankovich+2019}.}  
	\label{fig:winds} 
\end{figure*}

The fact that below the water condensation region in all four planets and the methane condensation region in Uranus and Neptune the temperature structure may show large temperature fluctuations (see Table~\ref{tab:clouds}) however invites a reexamination of this crucial hypothesis. This is further strengthened by the Juno observations of ammonia, and the fact that lightning, a tracer of water storm activity, is highly latitudinally dependent \citep{Brown+2018}.  

This should be addressed by a measurement of temperature as a function of latitude below the methane condensation region (preferably down to $\sim 100$\,bar similarly to the Juno MWR measurements).

\subsection{Rotation and magnetic field}


Giant planets rotate rapidly and have a strong differential rotation, Jupiter and Saturn showing equatorial superrotation, Uranus and Neptune subrotation (see Fig.~\ref{fig:winds}). Although mechanisms responsible for transferring angular momentum to or from the equator have been identified \citep[e.g.][]{Vasavada+Showman2005, Liu+Schneider2010}, the reason for the direction of the flow and its magnitude remain largely unknown. 

It may be not as well known that the rotation period of the deep interior in giant planets is not well known, except in Jupiter for which the planet's tilted magnetic field yields a well-defined periodic signal. In the case of Saturn the coincidence (to the level of accuracy of the measurements) between the spin and magnetic axes prevents this direct determination. The Voyager values for Saturn, but also for Uranus and Neptune were instead based on a period identified in the radio planetary signal that was close to the inferred deep rotation of the planet. The Cassini mission has since shown that this period was offset by several minutes, corresponding in Fig.~\ref{fig:winds} to an offset in the wind profiles by about 100 m/s \citep{Read+2009,Helled+2015,Mankovich+2019}. One of these methods, shape minimization, was also applied to Uranus and Neptune and predicts an offset with system III rotation that is equivalent to about +50 m/s and -100 m/s, respectively \citep{Helled+2010}. (Note that the true wind profile should be rederived to account for this effect instead of just applying an offset, but this is minor.) 

Interestingly, two features close to Neptune's South pole, the South Polar Wave and the South Polar Feature have had an extraordinary rotational stability for $\sim 20$\,years, with a period of $15.9663\pm 0.0002\,$h, close to Neptune's Voyager radio rotational period of $16.108\pm0.006\,$h \citep{Karkoschka2011a}. By comparison, the solid-body rotation period derived by shape minimization is $\sim 17.46\,$h \citep{Helled+2010}. Are we mislead by the shape minimization method or is there a surprising offset between the rotation of Neptune's polar region and its deep interior rotation? 

In any case, this calls for a direct determination of the shape of these planets, which directly affect the constraints that we can derive on their internal structure and composition. 

Of course, Uranus and Neptune have surprisingly complex magnetic fields \citep{Ness+1986, Ness+1989} that may be generated in a thin shell \citep{Stanley+Bloxham2006} or a thick shell \citep{Soderlund+2013}. Further observations of these planet's dynamos at high resolution would be invaluable to understand how these powerful magnetic fields are generated and how they couple to the interior structure that may be derived.

\subsection{Gravitational moments}

In any case, constraints on the interior structure and composition rely on accurate gravitational moments \citep[e.g.][]{Guillot2005}. Table~\ref{tab:grav} shows  the impressive improvement by about 2 orders of magnitude between measurements acquired from flyby measurements from Voyager spacecrafts  \citep[see][and references therein]{Guillot+Gautier2015} and when accounting for measurements by Juno \citep{Iess+2018} and Cassini \citep{Iess+2019}, on the basis of measurements by spacecrafts close-in and on polar orbits.

\begin{table}
\caption{Relative accuracies of gravitational moments}\label{tab:grav}
\begin{tabular}{lccc} \hline\hline
& $\sigma_{J_2}/J_2$ & $\sigma_{J_4}/J_4$ & $\sigma_{J_6}/J_6$ \\ \hline
Jupiter: Voyager & $6\times 10^{-5}$ & $9\times 10^{-3}$ & $0.6$  \\
\hphantom{Jupiter:} Juno & $9\times 10^{-7}$ & $7\times 10^{-6}$ & $3\times 10^{-4}$ \\
Saturn: Voyager & $2\times 10^{-5}$ & $3\times 10^{-3}$ & $0.1$ \\
\hphantom{Saturn:} Cassini & $2\times 10^{-6}$ & $4\times 10^{-5}$ & $1\times 10^{-3}$ \\
Uranus: Voyager & $9\times 10^{-4}$ & $9\times 10^{-3}$ & $-$ \\
Neptune: Voyager & $1\times 10^{-3}$ & $1\times 10^{-2}$ & $-$ \\ \hline\hline
\multicolumn{4}{l}{\parbox{\linewidth}{See \cite{Guillot+Gautier2015} and references therein for Voyager values, \cite{Iess+2018} for Juno values and \cite{Iess+2019} for Cassini values.}}
\end{tabular}
\end{table}

The increased accuracy in the gravitational moments of Jupiter and Saturn led to the determination for the first time of the depth of the zonal flows of these planets, i.e., about 3000\,km in Jupiter \citep{Kaspi+2018, Guillot+2018} and about 9000\,km in Saturn \citep{Iess+2019, Galanti+2019}. This value corresponds to a region in which hydrogen conductivity has increased to a level of about 1\,S/m (about the value in the Earth's oceans) and any significant differential rotation would yield the dissipation of an internal heat flux larger than the intrinsic heat flux of Jupiter \citep{Cao+Stevenson2017, Wicht+2019}. 

The gravitational moments of Uranus and Neptune are uncertain. As seen in Table~\ref{tab:grav}, only $J_2$ and $J_4$ have been determined. Still, an upper limit to the depth of the observed atmospheric zonal flows has been determined. It is about 1000\,km \citep{Kaspi+2013}. A precise value should be within easy reach of new gravity field measurements from an orbiter of these planets. It would provide invaluable information to understand the link between the planetary atmospheres and interior.

\subsection{Seismology}

Seismology is of course an invaluable tool to probe planetary interiors, as already demonstrated for Saturn \citep{Fuller2014, Mankovich+2019}. Jupiter is probably also seismically active \citep{Gaulme+2011}, although this is yet to be confirmed. The question of the excitation of the oscillations remains, particularly when extended to Uranus and Neptune \citep{Markham+Stevenson2018}. 

Unfortunately, only Saturn has large rings that are a perfect amplifier of planetary normal modes. For the other giant planets, one must rely on continuous observations of the planetary disk for at least days to detect waves that have amplitudes of at most a few tens of cm/s. Ground based observations at Jupiter with a Doppler imager appear promising - yielding already the direct measurement of zonal winds, with the ability to potentially measure meridional and vertical speeds \citep{Gonccalves+2019}. Given the potential of the method, the application to a spacecraft at Uranus or Neptune should be studied.

\subsection{Interiors of Uranus and Neptune}

\begin{figure*}
	\includegraphics[width=\linewidth]{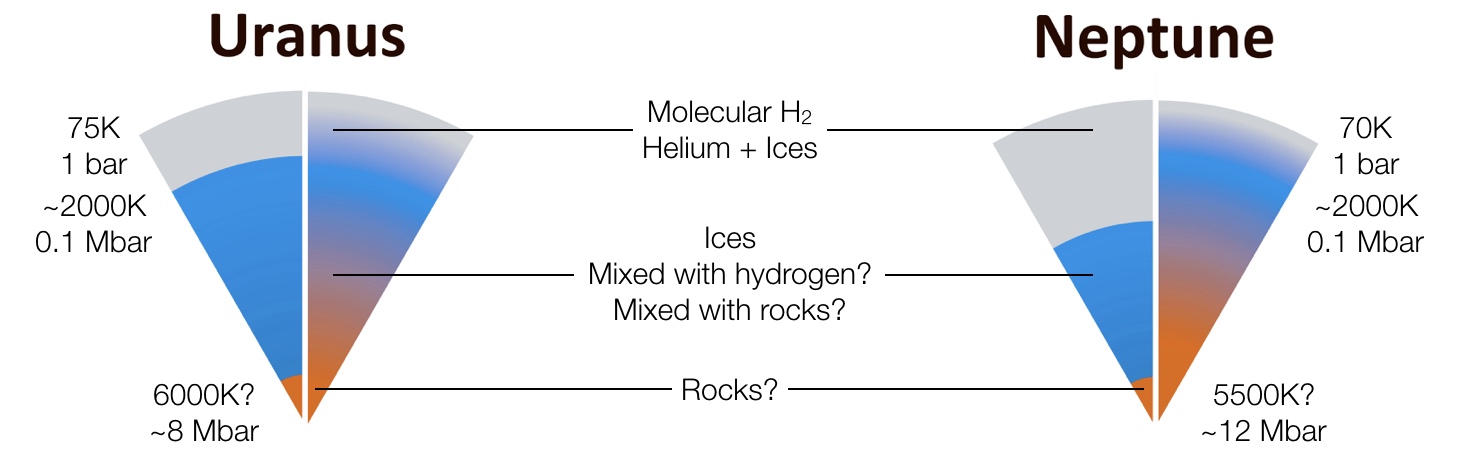} 
	\caption{Possible interiors of Uranus and Neptune \citep[adapted from][]{Helled+Guillot2018, Nettelmann+2013}}  
	\label{fig:interiors} 
\end{figure*}

We know that Uranus and Neptune have an envelope of hydrogen and helium of about 1 to 4 Earth masses and that their interior is denser \citep[e.g.][]{Nettelmann+2013,Helled+Guillot2018}. However, given the uncertainties on the gravitational moments, rotation rate, interior temperature profile, our lack of knowledge of heat transport in the presence of compositional gradients, we should be extremely cautious with the constraints that can be derived. For example, when assuming three-layer made of hydrogen and helium, ices and rocks and an adiabatic structure, one generally derives an ice to rock ratio that is significantly higher than the solar value \citep[e.g.][]{Podolak+1991, Hubbard+1995, Nettelmann+2013}, hence perhaps the term "ice giants" for these planets. However, the interior may not be adiabatic due to an inhibition of convection at transition layers, therefore leading to a retention of the interior heat \citep[as already discussed by ][]{Hubbard+1995}. Also, the high temperature early on during the formation of the planet may have prevented a differentiation of the elements into clean layers. When accounting for these uncertainties, it is likely that one can accommodate a large variety of ice to rock ratios, including the solar one. 

Figure~\ref{fig:interiors} shows possible structures for the interiors of Uranus and Neptune and highlights some of the uncertainties in present models. The interior temperature is particularly uncertain because of the unknown behavior between layers. Constraints on its value will probably be best derived from models of their formation. 

\begin{figure} 
	\includegraphics[width=\linewidth]{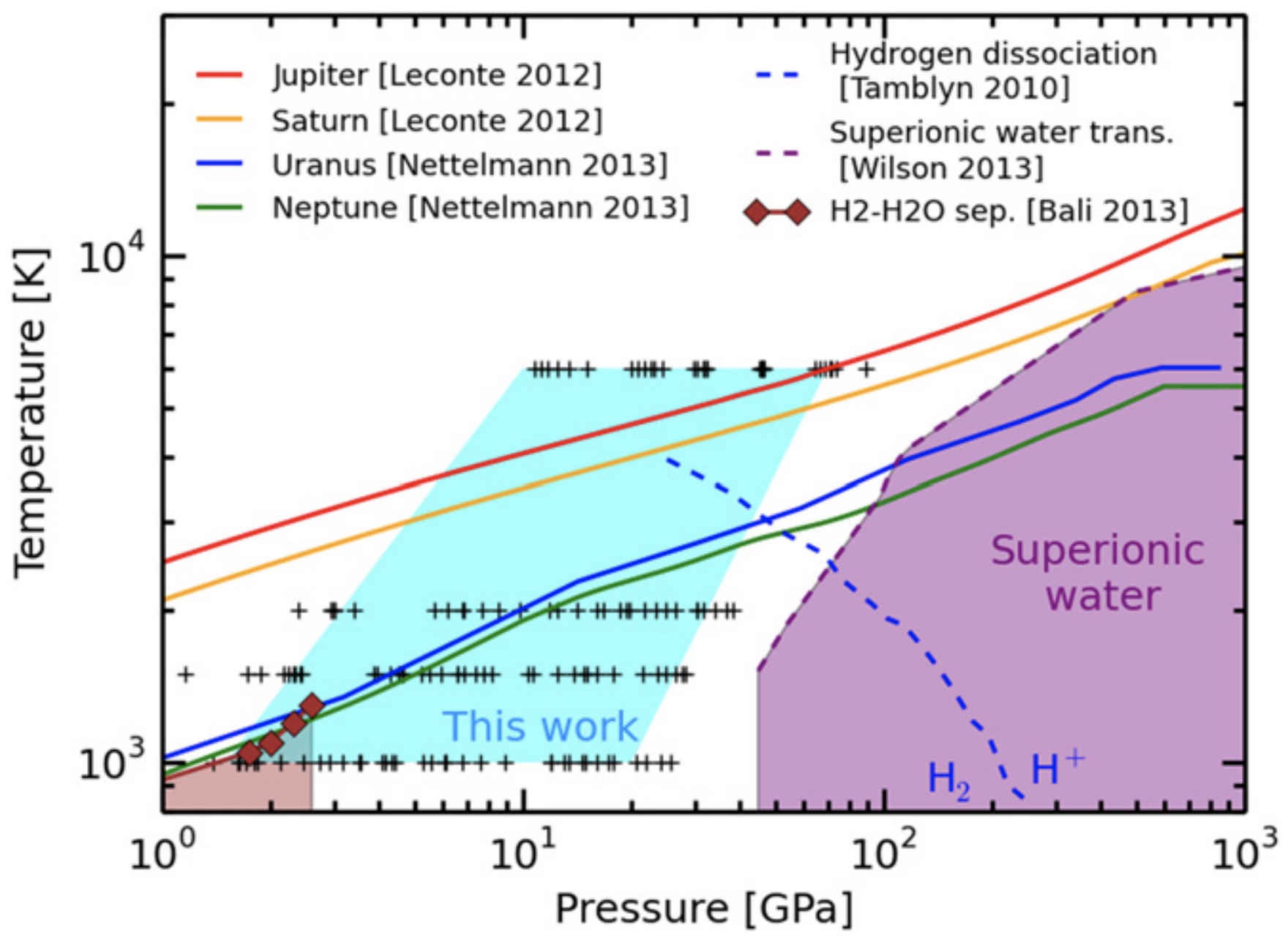} 
	\caption{Pressure-temperature diagram with the predicted interior profiles for the solar giant planets and the expected phase transitions of hydrogen and water. A phase separation between H$_2$ and H$_2$O at relatively low pressures (around 2 GPa, or 20 kbar) was seen in high-pressure experiments \citep{Bali+2013} and is shown in light brown. However, this phase separation was not observed in the numerical simulations of \cite{Soubiran+Militzer2015}, who probed conditions in the cyan area. \citep[from][]{Soubiran+Militzer2015}.}  
	\label{fig:soubiran} 
\end{figure}

Interestingly, the gravitational moments (say, $J_4$ to $J_8$) probe a region of the planet that essentially corresponds to the outer 1/3 in radius \citep[e.g.][]{Guillot2005}. This zone probably sees many important transitions, including the region of water condensation (see Fig.~\ref{fig:clouds}). Slightly deeper, but at moderate pressures and temperatures (~20kbar and ~1200\,K) it is possible that water becomes insoluble in hydrogen \citep{Bali+2013}. This would lead to the formation of a water ocean \citep[see][]{Bailey+Stevenson2015}. The possibility of such a phase transition was however not found in numerical simulations \citep[][and Fig.~\ref{fig:soubiran}]{Soubiran+Militzer2015}.  

At deeper levels, Fig.~\ref{fig:soubiran} shows that we first cross a first-order transition of metallic hydrogen near 0.4 Mbar. However, according to models \citep[e.g.][]{Nettelmann+2013}, we should not be dominated by hydrogen at this depth. Around 1\,Mbar, we should cross the transition at which water becomes superionic \citep{Cavazzoni+1999, French+2009, Wilson+2013}, creating a solid lattice of oxygen ions surrounding by a sea of free hydrogen atoms. The transition to such a solid phase should have profound consequences on the cooling of the planets. 

Further investigation of the interiors of Uranus and Neptune will require better gravitational moments. Without seismology it is likely that solutions will remain highly degenerate. We should in any case obtain constraints on the mass of hydrogen and helium present in the planets. However, constraining the ice to rock ratio will remain model dependent. We estimate however, that probing the methane condensation region will be useful to apply to other compositional gradients in the planets' interiors, building confidence in the interior models constraints.

\section{Formation, evolution and relation to exoplanets}

Several planetary embryos of sizes comparable to those of Uranus and Neptune may have existed even when Jupiter and Saturn had already reached their final mass \citep[see][]{Izidoro+2015}. It is also likely that planets of this mass abound in the Universe \citep{Fulton+2017}. But rather than the mass of these objects, what is key to such a mission is the fact that it applies to all planets with hydrogen atmospheres, particularly those for which we expect molecular weight gradients to be an important part of their structure and evolution, such as super-Earths with hydrogen rich atmospheres \citep[e.g.][]{Miller-Ricci+2009, Ikoma+Hori2012}. Knowing how heat and chemicals are transported in Uranus and Neptune's atmospheres will provide us with the tools to interpret future spectra of spatially unresolved exoplanets with hydrogen atmospheres.  

In the early stages, planetary embryos should possess a hydrogen atmosphere that is polluted with heavy elements. These elements, in particular water, ammonia and methane are expected to have a large impact on the cooling and therefore final properties of these forming planets \citep{Kurosaki+Ikoma2017}. Understanding how heat is transported in these atmospheres requires comparisons to direct measurements in Uranus and Neptune. 

Of course, the evolution of Uranus and Neptune themselves, with Uranus having an order of magnitude smaller intrinsic heat flux than Neptune \citep{Pearl+Conrath1991} remains a mystery. We do not have the solution, but it certainly requires a complete understanding of heat transfer in these planet's atmospheres. Being able to better spot the difference in internal structures of Uranus and Neptune, as determined from the measurement of their gravitational moments and magnetic fields will be crucial as well. 

Finally, some measurements performed in Uranus and Neptune can help reconstruct the history of the formation of the Solar System. Noble gases are particularly important because they could only be trapped at very low temperatures in the protosolar disk. Their abundance in the atmospheres of Uranus and Neptune compared to that in Jupiter would be an essential piece of the puzzle to determine e.g. whether photoevaporation in the late solar system or clathrate formation may have taken place \citep{Guillot+Hueso2006, Monga+Desch2015, Mousis+2009}.

\section{Objectives of a mission to Uranus or Neptune}

It is beyond the scope of the present paper to propose a mission scenario including instruments and a timeline. I will merely provide what I believe should be a few of the objectives of such a mission and when possible indicate how these objectives may be achieved. 

First, I cannot find, from a scientific point of view, a reason to clearly prefer one planet over the other. Uranus's tilt is interesting because it offers naturally an orbit that is pole on, and of course because it probably implies that the planet underwent a giant impact. Neptune appears to be more convectively active which makes it well suited for a study of moist convection. However Uranus may offer more quiescent regions to send a probe, while showing signs of regular convective activity. A decision to choose one planet or the other will have to be based on celestial dynamics or programmatic arguments. Of course, having the possibility to examine both planets would enable an extremely fruitful comparison. 

The three main goals of such a mission are (1) to understand transport processes in hydrogen atmospheres, (2) to constrain the structure, interior composition and dynamo of Uranus and/or Neptune and (3) to provide keys to understand the origin of the solar system. They are detailed afterwards with some of the essential measurements required. 

\subsubsection*{Understanding transport processes in hydrogen atmospheres}
Our lack of understanding of the transport of heat and of chemical species in the presence of compositional gradients and condensates limits our ability to model the interior structure and evolution of planets with hydrogen atmospheres, including all gas giants, but also exoplanets with hydrogen atmospheres or forming planetary embryos. This implies several measurements:
\begin{itemize}
\item Map temperature and methane abundance in the 1-5 bar region, as a function of latitude
\item Probe the deep atmosphere by measuring the brightness temperature as a function of latitude down to levels of $\sim 100$ bar 
\item Obtain a direct temperature profile that can be used to lift the degeneracies in other measurements. 
\item Monitor variability, storms, during several years
\item Map the concentrations of disequilibrium species and ortho/para hydrogen fraction as a function of latitude. 
\end{itemize}  

Mapping temperature and methane abundance at modest depths could be performed by an imaging spectrograph. This could also be used for the concentration of disequilibrium species and to the ortho/para hydrogen fraction with adequate, far-infrared, capabilities. Alternatively, it is possible that dedicated instruments may allow monitoring pre-determined species. 

Measurements in the deep atmosphere could be done by a microwave spectrometer. The very low abundance of ammonia inferred from the presence of an H$_2$S cloud \citep{Irwin+2018, Irwin+2019b} appears to be favorable for probing deeper into the atmosphere. However, given that the brightness temperature measurements are a function of both the physical temperature and the absorption coefficients a separate measurement of temperature at a well-defined location would lift the degeneracy. 

In order to get constraints on temperature profiles at several locations, we could rely on radio occultation measurements. It is important to realize that because of the relatively low atmospheric opacities at radio wavelengths, the Voyager radio occultations were able to reach a level of 2.3 bar in Uranus and 6.3 bar in Neptune \citep{Lindal1992}. However, radio occultations, however important and interesting, provide a constrain on the refractivity profile, itself a function of temperature and mean molecular weight (i.e., abundance of the condensing species). A direct measurement is essential to truly lift this degeneracy. 

With the inclusion of a probe, we would have this direct in situ measurement, both of temperature and condensate abundance. This requires however that it corresponds to a well-known location which is also observed by the instruments on the orbiter (spectrograph, MWR) at about the same time. Given the time- and space-variability of the atmospheres (see Fig.~\ref{fig:globes}), it would be important to choose a probe site that is reasonably stable. One possibility to be explored would be to add a few mini-probes to just measure the temperature-pressure profiles at several locations separated by a few to a few 100km. These mini-probes would provide an estimate of the small-scale variability of the temperature field, a very important aspect of the problem. 

Of course, both planets have an activity that changes on multiple timescales. As inferred from Fig.~\ref{fig:globes}, monitoring the atmospheric evolution and the development of storms and complex weather systems will require observations for several years.  
Both for the imaging spectrograph and the microwave radiometer, the requirement to observe different latitudes implies that the spacecraft should be at least part of the time away from the equatorial plane of the planet.

\subsubsection*{Constraining the structure, interior composition and dynamo of Uranus and Neptune}
Given the large total mass and decisive contribution of giant planets to shape the solar system that we know today, it is essential that we progress on constraining their interior structure and bulk composition. Part of the effort is to better understand transport processes, thus yielding a better understanding of the temperature structure of the planets. In order to improve model constraints on the mass of hydrogen and helium present, the ice to rock ratio and the structure of the interior, we need to:
 \begin{itemize}
\item Measure the interior rotation rate from the magnetic field rotation. 
\item Measure gravity field accurately (at least to $J_8$) 
\item Measure magnetic field accurately
\item Measure energy balance and interior heat flux
\end{itemize}  

Experience from Juno and Cassini indicates that the measurements needed will be best obtained with a spacecraft on a near polar orbit that comes very close to the surface of the planet. Given the complexity of the interior of Uranus and Neptune it would be important to look for small tesseral components of the gravity field which could probably trace slow-scale evolution of inhomogeneous structures in the planets' interiors. Similarly, the detection of secular variations of the magnetic field would be highly significant and could constrain mission duration. 

The determination of the planets' global heat balance and intrinsic luminosity is essential to understand the interior structure and evolution. This is particularly true for Uranus for which only an upper limit on the intrinsic luminosity is known. 

In addition, observations with a Doppler imager should lead to the detection of global oscillations of the planet. The identification of seismic modes would then provide a very powerful tool to detect discontinuities and provide constraints on the interior structure and composition.

\subsubsection*{Providing keys to understand the origin of the solar system}
Constraints on the interior compositions of Uranus and Neptune will naturally provide essential information to complete the inventory of the Solar System and understand its origins. It is also important to realize that the information gained from the examination of the methane condensation region should be applicable when modeling the water condensation regions in Jupiter and Saturn, and for exoplanets. Thus, such a mission would have wide implications to better understand planet formation in general. 

Several other key measurements would put Uranus and Neptune in context with other objects in the Solar System and provide more direct information on its early formation:
 \begin{itemize}
\item Measure noble gases composition (including helium) 
\item Measure isotopic composition of several key species including Deuterium, He, C, Ne, Ar
\end{itemize}  

The noble gas composition is to be compared to similar measurements in Jupiter by the Galileo probe to determine how these where delivered to the giant planets. In particular, a measurement of a constant enrichment over solar in Ar, Kr and Xe would be a strong indication in favor of photoevaporation in the protosolar disk.  

Determining the helium to hydrogen ratio directly would be essential: Although we expect it to be equal to the primordial protosolar value, we do not know for sure. In Jupiter and Saturn, helium sinks to the interior of the planet, something that is not expected for Uranus and Neptune which should not have an important metallic hydrogen reservoir. This could thus be a way to determine precisely the protosolar helium to hydrogen ratio, an essential ingredient for models of Jupiter and Saturn and for the Sun. Alternatively, if it turns out that this value differs, we will have to strongly revise our models of formation of the solar system and/or of giant planet interiors. 

The measurements of isotopic compositions of several key species (not necessarily limited to the above list) would give the possibility to place Uranus and Neptune in context with other objects in the solar system including Jupiter but also terrestrial planets, comets and asteroids. 

These measurements are best done with a probe. Since they are not tied to condensation issues or chemical reactions they can be performed at one location and generally with a relatively shallow probe.

\subsubsection*{Looking for surprises}
One of the goals of such an ambitious mission is also to expect the unexpected: As we have seen, our understanding of the mechanisms at play in these planets is limited. Jupiter and Saturn are, in comparison, relatively simple when we consider the multiple possibilities in terms of phase changes in Uranus and Neptune. The suite of instruments carried by the spacecraft (not necessarily limited to the ones described here) should give us the possibility to gain further understanding from the surprises that we should expect.   

An important aspect will also be a simultaneous monitoring of these planets to look for storms or changes in the atmosphere. With new adaptive optical instruments becoming available, it is expected that amateur astronomers could also provide useful contribution and participate to the mission.

\section{Conclusion}

Uranus and Neptune hold some of the keys to understand planets with hydrogen atmospheres, finalize the inventory of the Solar System and infer the history of its formation. A dual mission with an orbiter and a probe reaching all the objectives described in this proposal would be best achieved through an international collaboration.  It will be a much awaited milestone in the exploration of our Solar System and will provide the tools needed for a the interpretation of observations of planets in our Galaxy.


\printbibliography[title={Bibliography}] 


\end{document}